\documentclass[preprint]{revtex4}

\usepackage{graphicx}
\usepackage{dcolumn}
\usepackage{amsmath}

\begin{document}

\title{Peltier ac calorimeter}

\author{D. H. JUNG, I. K. MOON, and Y. H. JEONG}
\email[YHJ wishes to dedicate this paper to Dr. G. H\"ohne.]{}
\affiliation{Department of Physics and electron
Spin Science Center, Pohang University of Science and Technology,\\
Pohang, Kyungbuk 790-784, S.  Korea\vskip 1cm}

\begin{abstract}
{A new ac calorimeter, utilizing the Peltier effect of a
thermocouple junction as an ac power source, is described. This
Peltier ac calorimeter allows to measure the absolute value of
heat capacity of small solid samples with sub-milligrams of mass.
The calorimeter can also be used as a dynamic one with a dynamic
range of several decades at low frequencies.}
\end{abstract}

\maketitle

\section{INTRODUCTION}

At the turn of the 20th century  studies of heat capacity of
solids at low temperatures played an important role in revealing
the quantum character of Nature.~\cite{nature} This example
vividly illustrates the power of heat capacity measurements in
physics. As a matter of fact measurements of heat capacity reveal
so great a deal of information about matter that calorimetry has
become an indispensable tool for modern day research in chemistry,
physics, materials science, and biology. Unfortunately, however,
calorimetry is a relatively insensitive method, and it is
particularly difficult to obtain an accurate absolute value of
heat capacity of samples with minute masses. Since new materials
with interesting physical properties are usually not synthesized
in quantity, it is of extreme necessity for the advance of
materials science or condensed matter physics to secure a
convenient means to measure absolute heat capacity of
sub-milligram samples.

On another front, the generalization of calorimetry into the
dynamic regime has attracted wide attention in recent
years.~\cite{lahnwitz} Although measurement of a thermodynamic
quantity is the usual notion that is tied to calorimetry, it is
possible to go beyond this traditional understanding and
generalize heat capacity as a dynamic quantity. The concept of
dynamic heat capacity appears natural if one recalls that static
thermodynamic quantities are time-averaged (or ensemble-averaged).
In other words, they are static not because they do not change in
time, but because they change too rapidly on the experimental time
scale. Then, suppose that a system contains a dynamic process
relaxing with a characteristic time which lies within our
experimental time window, this will result in a time-dependent (or
frequency-dependent) heat capacity depending on the time scale of
measurements. Conversely, measurements of dynamic heat capacity of
condensed matter would provide insights which may not be available
to other dynamic probes. Thus, the development of a convenient
dynamic calorimeter for solid samples appears to be of great
necessity.

In this paper, we describe a new type of the ac calorimeter,
termed {\it Peltier ac calorimeter} (PAC), which fills both needs
described above; PAC is not only a microcalorimeter capable of
measuring heat capacity of sub-milligram samples but also a
dynamic calorimeter with wide dynamic range. For the heat capacity
measurements of a minute sample adiabatic
calorimetry~\cite{nernst}  does not appear to be suitable due to
the so-called addenda problem; in other words, calorimetry
requires indispensable addenda (heater and sensor) to be put on a
sample and the mass of the addenda may even be greater than that
of the sample in the case of minute samples. Although one may
expect the same kind of problem in ac calorimetry,~\cite{sullivan}
we have devised a way of avoiding the addenda problem in a new ac
calorimeter by utilizing the Peltier effect of extremely thin
thermocouple wires. It is also obvious that an ac calorimeter has
a potential of being a dynamic one.

\section{Principle of the Peltier ac calorimeter}
Suppose that a voltage difference $\Delta V$ and/or a temperature
difference $\Delta T$ exist across a metallic wire, the electric
current $I$ and the heat current $P$ through the wire can be
expressed as~\cite{callen}
%%%%%%%%%%%%%%%%%%%%%%%%%%%%%%%%%%%%%%%%%%%%%%%%%%%%%%%%%%%%%%%%%%%%%%%%%%
\begin{equation}
 I = -(\Delta V + S\Delta T)/R\
 \end{equation}
 \begin{equation}
   P =
\Pi I - K\Delta T,
 \end{equation}
%%%%%%%%%%%%%%%%%%%%%%%%%%%%%%%%%%%%%%%%%%%%%%%%%%%%%%%%%%%%%%%%%%%%%%%%%%
where $R$ is the resistance, $S$ the thermoelectric power, $\Pi$
the Peltier coefficient, and $K$ the thermal conductance. The
thermoelectric power and the Peltier coefficient are related by
$\Pi = TS$. Now if an electric current is run through a
thermocouple, consisting of two distinct metal wires, under an
isothermal condition, then the junction acts as either a heat sink
or a heat source depending on the current direction. The heat
current $P$ caused by this Peltier effect is directly proportional
to $I$. If an ac electric current at an angular frequency
$\omega$, $I(t) = I_0\exp(i\omega t)$, is applied to a
thermocouple under an isothermal condition, an ac power
oscillation at the same frequency will be induced at the junction
by the Peltier effect. The amplitude $P_0$ of the ac power is
equal to $P_0\,=\,T\,\Delta S\,I_0$; $P_0$ can be accurately
determined, since $\Delta S$ is well tabulated for thermocouples
and the ac current can be measured with high precision. Thus it is
evident that the Peltier effect of a thermocouple junction can be
utilized as a power source for ac calorimetry.

Various ac calorimetric techniques with non-contact energy sources
such as chopped light or light emitting diode were developed
previously.~\cite{hatta,cal_LED} However, it should be pointed out
that these power sources can only supply heat and cannot act as a
heat sink. As a consequence of this, the average temperature of a
sample is always above that of the heat bath in these traditional
ac calorimetric methods. (This so-called dc shift is equal to the
average dc power divided by the thermal conductance between the
sample and the bath.) In addition, it is not easy in traditional
ac calorimetry to determine an absolute value of heat capacity due
to inaccuracy in the determination of input power and heat leak.
On the other hand, the PAC which utilizes the Peltier effect as an
ac power source is free from these difficulties: first, it is
capable of both heating and cooling. This means that there is no
dc shift in temperature, and lack of the dc shift in turn has an
important implication on the working frequency range of the ac
calorimeter. (See below.) Second, the power generated at a
thermocouple junction can be measured with high accuracy. Third,
the mass of the thermocouple junction attached to the sample is
entirely negligible in most situations. All these factors make the
PAC superior to previous methods in that the experimental setup is
simple, it directly yields absolute values of heat capacity of
sub-milligram samples, and it may be used as a dynamic
calorimeter.

\section{Implementation of the principle}
The implementation of the principle described in the previous
section is rather straightforward; the schematic diagram and the
photograph of the Peltier ac calorimeter we constructed is shown
in Fig.~1. We made thermocouple junctions by spot-welding with
chromel and constantan wires of 25 $\rm\mu m$ (or 12 $\rm\mu m$)
in diameter. A couple of thermocouple junctions (TC1) were
connected via copper wires to a function generator supplying an ac
electric current, which was measured by a digital ammeter with 1
nA sensitivity. Using a very small amount of GE 7031 varnish for
electrical insulation and good thermal contacts, one of the
junctions was attached to one side of a sample (typically of
linear size less than 1 mm) and the other to a copper block (heat
bath). The heat bath was then attached to a closed-cycle He
refrigerator and its temperature was controlled within $\pm$ 2 mK
stability  in the range of 15--420 K. A digital voltmeter was used
to read the voltage difference across another thermocouple
junction (TC2) attached to the other side of the sample. From the
voltage readings, we could measure the sample temperature with the
sensitivity of 1.67 mK at 15 K and 0.14 mK at 420 K.

Since the mass of the thermocouple junctions and the varnish is
completely negligible compared to the sample mass even for
sub-milligram samples, the amplitude of the temperature
oscillation $\delta T_\omega$ at the imposed frequency (measured
by TC2) can be written as~\cite{hatta}
%%%%%%%%%%%%%%%%%%%%%%%%%%%%%%%%%%%%%%%%%%%%%%%%%%%%%%%%%%%%%%%%%%%%%%%%%%
\begin{equation}
\label{eq_accal} \delta T_\omega = \frac{P_0}{\omega C_p}\left(1 +
\frac{1}{\omega^2\tau_{ext}^2} + \omega^2\tau_{int}^2 +
\frac{2K_b}{3K_s}\right)^{-1/2},
\end{equation}
%%%%%%%%%%%%%%%%%%%%%%%%%%%%%%%%%%%%%%%%%%%%%%%%%%%%%%%%%%%%%%%%%%%%%%%%%%
where $C_p$ is the sample heat capacity, $\tau_{ext}$ the external
or sample-to-bath relaxation time, $\tau_{int}$ the internal
diffusion time in the sample, $K_b$ the thermal conductance of the
link between the sample and the bath, and $K_s$ the thermal
conductance of the sample. Since only the thin thermocouple wires
of diameter 12 or 25 $\mu$m provide paths for heat conduction from
the sample to the heat bath, $K_b$ is extremely small and
negligible compared to $K_s$. And the third term in the
parenthesis of Eq.~(\ref{eq_accal}) can be neglected. Thus, if one
can select the frequency range of $1/\tau_{ext} \ll \omega \ll
1/\tau_{int}$, heat capacity may be directly obtained from $C_p =
P_0/\omega\delta T_\omega$.

In addition to providing a means for measuring the absolute value
of heat capacity of minute solid samples,  the PAC possesses a
distinct ability to function as a dynamic calorimeter. This
capability of the PAC stems from the fact that $K_b$ is extremely
small as noted above (Heat conduction paths are provided by
microns thick thermocouple wires only.) and thus $\tau_{ext}$ is
exceedingly large. Large $\tau_{ext}$ then allows a wide frequency
range where the relationship $C_p = P_0/\omega\delta T_\omega$
holds. This situation is contrasted to that of traditional ac
calorimeters where one is required to have a reasonable size of
$K_b$, because otherwise the dc shift of the sample due to a dc
power would become prohibitively large. (Remember that the dc
shift is given by the dc power divided by $K_b$.) This size
requirement of $K_b$ then places a limit to the working frequency
range of the ac calorimeter by decreasing $\tau_{ext}$. Note that
the origin of this limitation of the traditional ac calorimeter
goes to the fact that the energy source is only capable of
heating, but not cooling. On the other hand, the energy source of
the PAC is able to both heat and cool, the average temperature of
the sample is the same as the bath temperature, and the PAC can
afford to have exceedingly small $K_b$ and long $\tau_{ext}$.

\section{PERFORMANCE of the Peltier ac calorimeter}

The performance of the PAC was tested with small pieces of
synthetic sapphire (${\rm \alpha}$-${\rm Al_2O_3}$), the standard
material designated by NIST.~\cite{cp_ref} We first examined the
induced temperature oscillation in TC2, in response to an
oscillating electric current in TC1, at 30 K, 150 K, and 320 K
with a test sample of mass 0.54 mg and dimension
1$\times$0.5$\times$0.3 mm$^3$. The frequency and amplitude of the
electric current were 0.25 Hz and 0.4 mA, respectively. From
Fig.~2(a) and (b), it is seen that nice temperature oscillations
at the applied frequency are obtained  at 150 K and 320 K.
However, Fig.~2(c) reveals that a significant amount of the second
harmonic appears at 30 K. It is readily clear that this second
harmonic originates from Joule heating in the thermocouple wires.
To precisely determine $\delta T_\omega$ which will yield heat
capacity, the raw data were fitted to a sinusoidal function
containing the fundamental and second harmonic terms. Fig.~2(d)
shows the power amplitude at the fundamental and second harmonics
as a function of temperature. The power amplitude at the
fundamental were converted from the measured $I_0$ using the table
for $\Delta S$.~\cite{table} The Joule heating part was obtained
from the fitting procedure. Above 50 K, the Joule heating effect
is completely negligible; however, there appears an increasing
amount of the second harmonic as temperature decreases below 50 K.
This is probably caused by the reduction in thermal resistance of
the thermocouple wires at low temperatures, and thus part of the
heat generated along the wires flows back toward the sample.
However, the existence of the second harmonic does not cause too
much problem in the heat capacity measurements, since the
governing equation for the present problem is linear and therefore
one needs only to measure the signal at the fundamental frequency.
Nevertheless, it is desirable to reduce the Joule heating as much
as possible to attain the sensitivity.

The dynamic characteristics of the PAC was then checked by
measuring the frequency dependence of $\delta T_\omega$ for the
test sample (0.54 mg) at a fixed amplitude of the applied current.
Fig.~3 is the plot of the results obtained at three temperatures.
It is seen from the figure that $\delta T_\omega$ is proportional
to $f^{-1}$ in the whole measured frequency range at high
temperatures (150 K and 320 K), and the figure clearly illustrates
that the PAC is indeed a dynamic calorimeter. The data obtained at
30 K, however, shows a large deviation from the $f^{-1}$ behavior
at frequencies below 0.13 Hz. This is caused by the fact that the
smallness of $K_b$ is compensated by a rapid decrease in heat
capacity at low temperatures, and the external relaxation time
$\tau_{ext}$ becomes short enough to be comparable to the
oscillation period at 0.13 Hz. This compensation effect is
unavoidable and the dynamic capability of the PAC is of limited
use below roughly liquid nitrogen temperature. It may be also
noted that the deviation from the $f^{-1}$ behavior is also
expected at high frequencies, since our power source is an
extremely local one and the internal diffusion time will interfere
above a certain frequency. This high cutoff should be
size-dependent, and present no problem for small samples.

In order to ascertain the capability of the PAC as a
microcalorimeter, we carried out the heat capacity measurements
for two test samples of ${\rm \alpha}$-${\rm Al_2O_3}$ with mass
0.54 mg and 2.25 mg. The measuring frequency was set at 0.25 Hz
and the measured temperature range was from 15 K to 420 K. It is
stressed that background subtraction, required in most
calorimetric methods, is not necessary for the PAC. Fig.~4 (a) is
the plot of the heat capacity data for two sapphire samples. The
two sets of data coincide very well and display excellent
reproducibility of the PAC. Also plotted in the figure is the
reference data ($C_p^{ref}$) of the same material.~\cite{cp_ref}
The agreement is again excellent in the whole temperature range
from 15 K to 420 K. In order to estimate the accuracy and
precision of the PAC, we plotted the residual heat capacity
values, {\it i.e.}, ($C_p-C_p^{ref}$) in Fig.~4(b). From the
figure, we estimate the absolute accuracy for heat capacity of a
sub-milligram sample to be $\pm 3$\% for the temperature range of
30--150 K and $\pm 1$\% for 150--420K. At temperatures below 30 K,
the absolute accuracy worsens due to very small values of heat
capacity of ${\rm \alpha}$-${\rm Al_2O_3}$ and becomes even more
than 10\%. However, the precision is better than 0.5\% in the
whole temperature range of 15--420 K.

\section{FUTURE OUTLOOK}
Having demonstrated the capability of the PAC as a
microcaloriemter and dynamic calorimeter, we briefly outline the
possible extensions of the PAC, which are currently under
development. First of all, it should be noted that the PAC would
function as a microcalorimeter even at Helium temperatures if
cryogenic thermocouple wires of Au-Fe or Cu-Fe are used. This
replacement of thermocouples would have an additional effect of
suppressing unwanted Joule heating. A more novel extension of the
PAC would be the Peltier thermal microscope, which would enable to
measure local thermophysical properties of matter at submicron
length scales. Here it is proposed that the tip for a atomic force
microscope is replaced by a thermocouple tip. An important feature
of our proposal is that the thermocouple tip here is not just a
temperature sensor, but it plays dual roles of a heater and
sensor. For this purpose, we have shown that a single junction can
indeed be used as both a heat source and  sensor
simultaneously.~\cite{jung} The successful development of the
Peltier thermal microscope would be an exciting and important
event for this so-called nano-age when the submicron local
thermophysical properties are in great demand.

\begin{acknowledgments}
{This work was supported by the electron Spin Science Center and
the BK21 program at POSTECH.}
\end{acknowledgments}

\begin{figure}
\includegraphics[height=6cm]{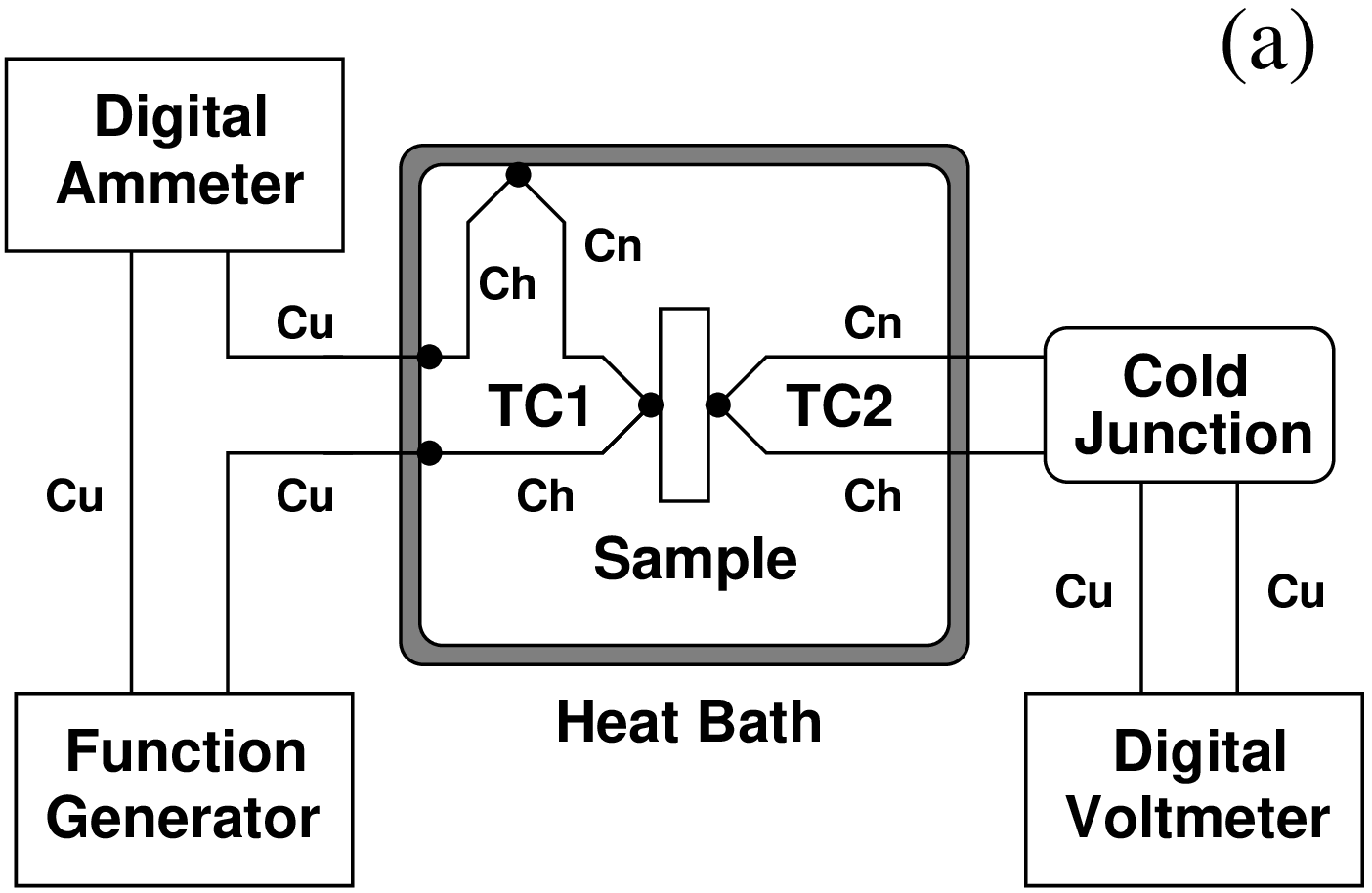}\vskip 1cm
\includegraphics[height=6cm]{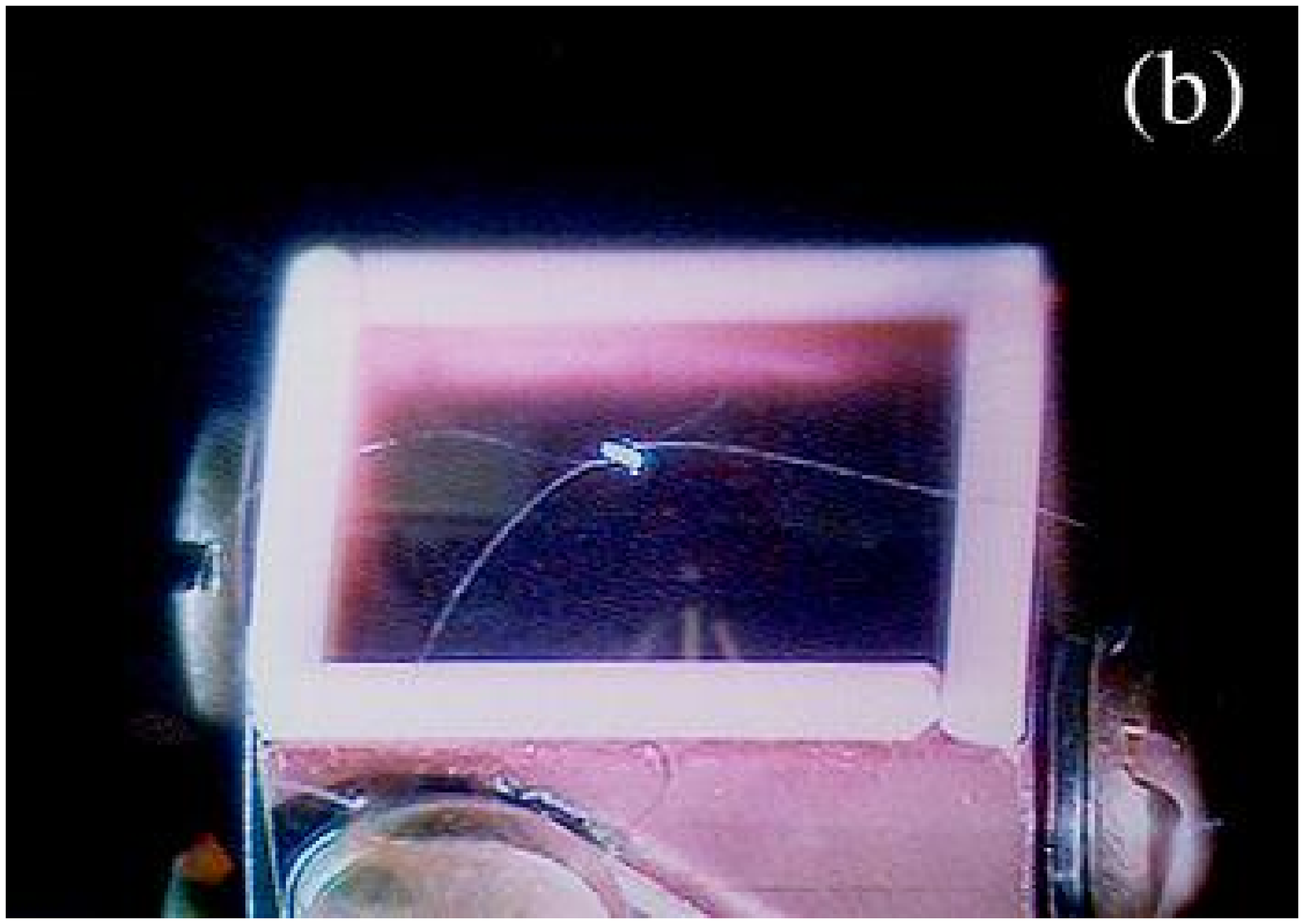}
\caption{(a) The schematic diagram of the Peltier ac calorimeter.
Ch, Cn, and Cu denote chromel, constantan, and copper wires,
respectively. Wires of either diameter 25 $\mu m$ or 12 $\mu m$
are used. A copper block plays a role of heat bath. A function
generator applies an oscillating current to Ch-Cn thermocouples
(TC1) in contact with the sample, and a digital voltmeter measures
an ensuing voltage oscillation from another thermocouple (TC2)
attached to the sample. (b) The photograph of the Peltier ac
calorimeter. The linear dimension of the sample is approximately 1
mm, and the diameter of the thermocouple wires is 25 $\mu m$.}
\end{figure}

\begin{figure}
\includegraphics[width=10cm]{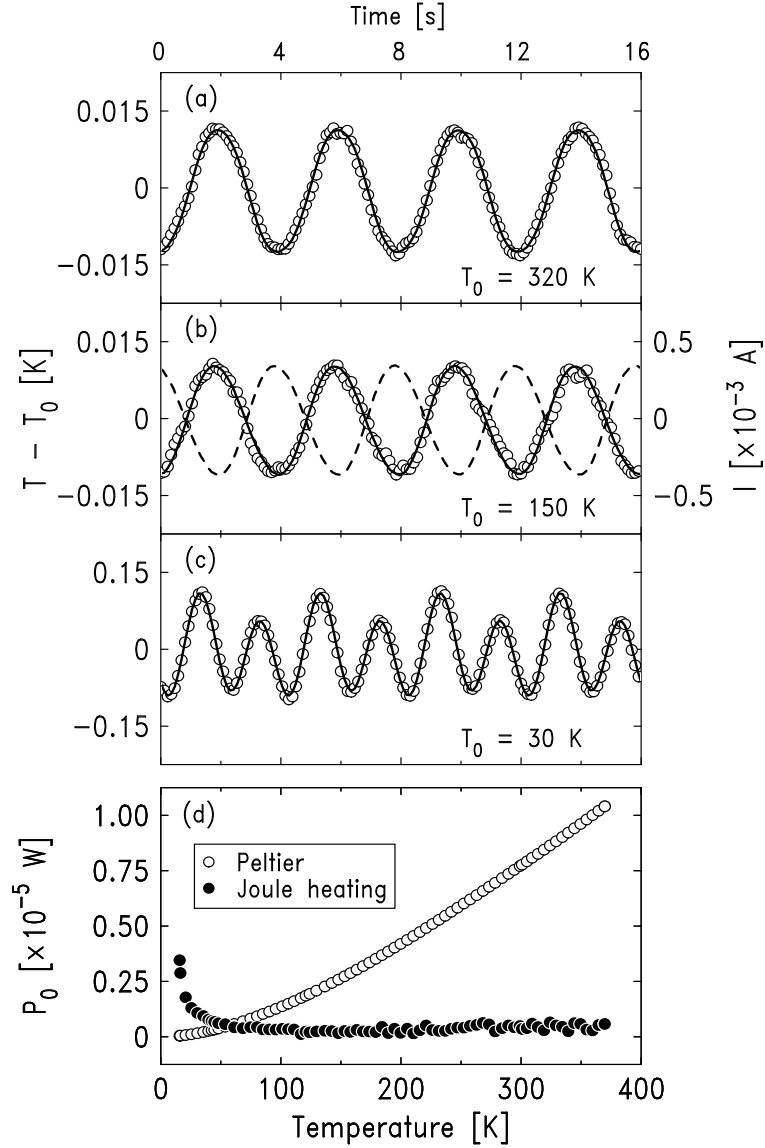}
\caption{The data of the sample temperature oscillation measured
at 30 K (a), 150 K (b), and 320 K (c). The test sample was a piece
of sapphire with mass 0.54 mg and dimension
1$\times$0.5$\times$0.3 mm$^3$. The broken line of (b) shows the
electric current oscillation. The solid lines indicate the fitting
results with the fundamental and second harmonics, and $T_0$
represents the dc component. (d) The amplitude of the power
oscillations at the fundamental (Peltier) and second harmonics
(Joule).}
\end{figure}

\begin{figure}
\includegraphics[width=10cm]{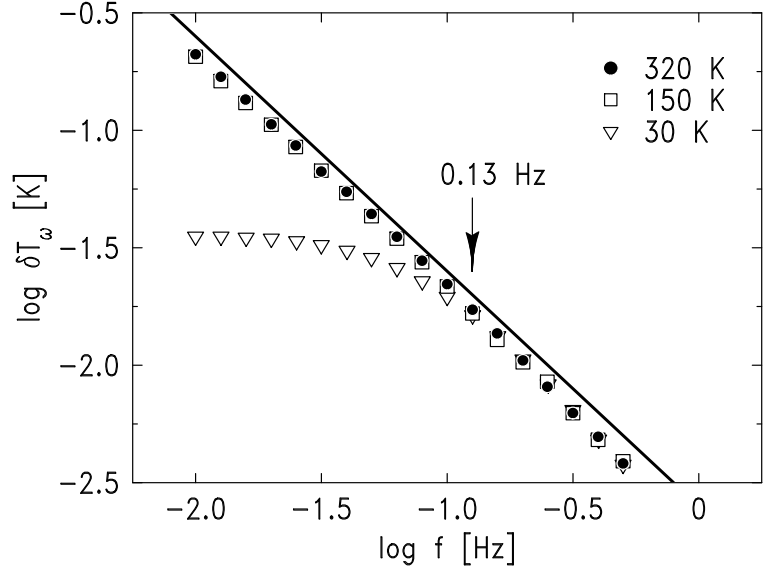}
\caption{The dynamic characteristics of the Peltier ac
calorimeter. The magnitude of temperature oscillation $\delta
T_\omega$ is plotted as a function of frequency. The thick solid
line represents the $f^{-1}$ behavior. Note that the the $f^{-1}$
law is well obeyed in the whole frequency range at high
temperatures; the PAC is able to function as a dynamic
calorimeter. The arrow indicates the low cutoff frequency below
which the deviation from the $f^{-1}$ behavior appears at low
temperatures.}
\end{figure}

\begin{figure}
\includegraphics[width=10cm]{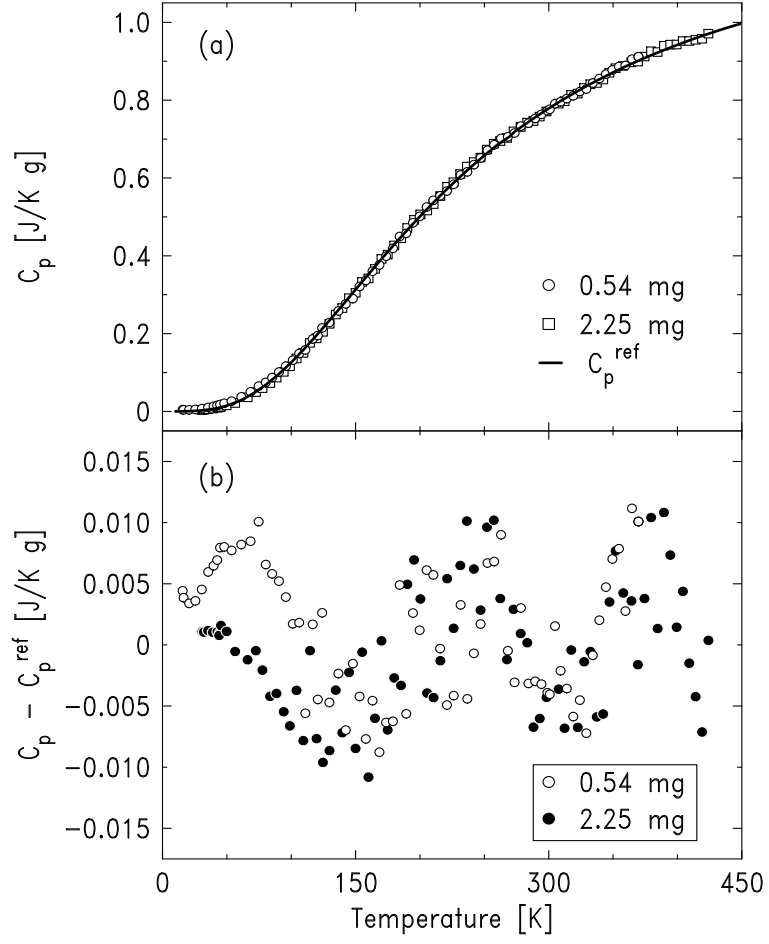}\caption{(a)
Specific heat capacity of the standard synthetic sapphire. $C_p$
of two samples with mass 0.54 mg and 2.25 mg is plotted. The solid
line represents reference values from Ref.~\cite{cp_ref}. (b) The
residual values, ($C_p-C_p^{ref}$), are plotted.}
\end{figure}

\end{document}